\documentclass[aip,pop,graphicx,reprint]{revtex4-1}

\usepackage{graphicx,amsmath,amsthm,amssymb,bm}

\draft 

\begin{document}


\title{Variational integration for ideal magnetohydrodynamics with built-in advection equations} 



\author{Yao Zhou}
\affiliation{Plasma Physics Laboratory, Princeton University, Princeton, New Jersey 08543, USA}
\author{Hong Qin}
\affiliation{Plasma Physics Laboratory, Princeton University, Princeton, New Jersey 08543, USA}
\affiliation{Department of Modern Physics, University of Science and Technology of China, Hefei, Anhui 230026, China}
\author{J.\,W.\,Burby}
\affiliation{Plasma Physics Laboratory, Princeton University, Princeton, New Jersey 08543, USA}
\author{A.\,Bhattacharjee}
\affiliation{Plasma Physics Laboratory, Princeton University, Princeton, New Jersey 08543, USA}


\date{\today}

\begin{abstract}
Newcomb's Lagrangian for ideal magnetohydrodynamics (MHD) in Lagrangian labeling is discretized using discrete exterior calculus. Variational integrators for ideal MHD are derived thereafter. Besides being symplectic and momentum-preserving, the schemes inherit built-in advection equations from Newcomb's formulation, and therefore avoid solving them and the accompanying error and dissipation. We implement the method in 2D and show that numerical reconnection does not take place when singular current sheets are present. We then apply it to studying the dynamics of the ideal coalescence instability with multiple islands. The relaxed equilibrium state with embedded current sheets is obtained numerically.
\end{abstract}


\maketitle 


\section{introduction}
Ideal magnetohydrodynamics (MHD) is a fundamental model in fusion, space, and astro plasma physics. It describes an ideal fluid with mass, entropy, and magnetic flux advected by the motion of the fluid elements. It is a Hamiltonian system with zero dissipation\cite{Newcomb,Morrison}. Newcomb \cite{Newcomb} first proposed the Lagrangian formulation for ideal MHD, in both Eulerian and Lagrangian labelings. In this formulation, the momentum equation follows from Hamilton's action principle, while the advection equations are built-in as constraints of motion. The theory of Euler-Poincar\`e reduction\cite{Holm} is a subsequent generalization of Newcomb's theory.

Ideal MHD is also a nonlinear system so complicated that numerical simulations are usually needed to solve it. However, most, if not all, known numerical methods for ideal MHD suffer from artificial dissipation, which limits the domain of applicability of the simulation results. Interestingly, numerical dissipation in ideal MHD simulations is characterized by more than just energy dissipation. A numerical scheme for ideal MHD can be energy conserving\cite{Liu} yet dissipative due to error introduced by solving the advection equations, which could lead to artifacts violating the frozen-in law. For example, field lines would typically break and reconnect along current sheets in ideal MHD simulations\cite{Gardiner}. Such numerical artifact will be referred to as ``numerical reconnection" and focused on in this paper, as it is directly observable as an indication for the dissipation introduced by solving the advection equations.

Variational integrators \cite{Marsden} are obtained from discretized Lagrangians as algorithms for simulating Hamiltonian systems. Such algorithms are useful for managing numerical dissipation because they naturally inherit many of the conservation laws of the continuous systems' . For a finite-dimensional system with a non-degenerate Lagrangian, variational integrators are known to be symplectic and momentum-preserving. For non-canonical systems\cite{Qin}, or infinite-dimensional systems such as electrodynamics\cite{Stern}, Vlasov-Maxwell system \cite{Squire}, and incompressible fluids\cite{Pavlov}, it becomes more challenging to discretize the Lagrangians while preserving the desired conservation laws. 

Nevertheless, there have been efforts to develop variational integrators for ideal MHD\cite{Kraus,Gawlik} that respect its conservation laws. In Ref.\,\onlinecite{Kraus}, a formal Lagrangian is used instead of the physical Lagrangian. In contrast to Newcomb's formulation, the frozen-in equation follows from the Euler-Lagrange equations associated with the formal Lagrangian. When singular current sheets are present, the method still suffers from numerical reconnection, despite conserving energy exactly. In Ref.\,\onlinecite{Gawlik}, discrete volume-preserving diffeomorphism groups are constructed on isolated cells such that non-local interactions between any two cells are included. As a result, non-holonomic constraints are required to localize the discrete velocity fields \cite{Pavlov}. However, it is not clear whether the method is still symplectic with such unphysical constraints. In addition, such a discretization cannot be applied to compressible MHD. 

Considering these imperfections of the previous attempts, we choose to develop variational integrators for ideal MHD along another path. One key guideline is to treat the equation of motion and the advection equations differently, respecting the nature of ideal MHD. Another is to construct the discrete Lagrangian on a discrete manifold with more structures than isolated cells. Therefore, we choose to discretize Newcomb's physical Lagrangian using discrete exterior calculus\cite{Desbrun,Hirani} (DEC). 

In this paper, we will perform this exercise, and then assess the viability of the resulting variational integrators. The discretization will be carried out on the Lagrangian in Lagrangian labeling. First, the Lagrangian will be spatially discretized using DEC, with built-in advection equations inherited from Newcomb's formulation. By so doing we will be using a finite-dimensional moving mesh to capture the infinite-dimensional MHD. The spatially discretized Lagrangian will have an $N$-body form, which we will then straightforwardly discretize in time and obtain variational integrators\cite{Marsden} to numerically solve for the motion of the mesh. We will show numerically that the method can handle singular current sheets without numerical reconnection. We will also simulate the relaxation of coalescence instability\cite{Longcope} and obtain the equilibrium with current sheets expected from ideal MHD theory, which previous studies failed to because of numerical reconnection. 

The proposed method brings about the first variational symplectic integrators for ideal compressible MHD. The two steps in discretization account for the two highlights of the integrators respectively. Variational temporal discretization makes the schemes symplectic and momentum-preserving. By building in the advection equations, we avoid solving them and the accompanying errors. Therefore, the method is very effective in problems where advection constraints are of dominant importance, outperforming any existing algorithms in terms of mitigating numerical reconnection. This suggests that it may be best suited to studying spontaneous current sheet formation in ideal MHD\cite{Parker}.

In a particle-based Lagrangian algorithm\cite{Rosswog}, solving the frozen-in equation is also avoided by advecting the so-called Euler potentials with the particles. Our discretization of the frozen-in equation as advection of discrete magnetic flux is more general, geometric, and physical. Simulating MHD with a moving mesh is not a new idea\cite{Pakmor}, but to our knowledge we are the first to discretize the ideal MHD Lagrangian and obtain variational integrators on it. Our schemes are vulnerable to mesh distortion like most moving-mesh methods\cite{Springel}, but can still be useful for certain problems. In fact, spontaneous current sheet formation has been studied with moving-mesh methods previously\cite{Craig}.

This paper is organized as follows. First, Newcomb's Lagrangian formulation for ideal MHD in Lagrangian labeling is briefly reviewed. Next, we introduce DEC to spatially discretize Newcomb's Lagrangian, and derive the variational integrators. Then we implement the method in 2D and show numerical results that artificial reconnection does not take place. In the end, the strengths and weaknesses of the method will be summarized and discussed.

\section{Lagrangian formulation for ideal MHD}
In this section, we review the Lagrangian formulation for ideal MHD first presented by Newcomb in Ref.\,\onlinecite{Newcomb}. He began with the Lagrangian for ideal MHD in Eulerian labeling $(\mathbf{x},t)$,
\begin{equation}
L(\mathbf{v},\rho,p,\mathbf{B})=\int\left(\frac{1}{2}\rho{v}^2-\frac{p}{\gamma-1}-\frac{1}{2}{B}^2\right)\mathrm{d}^3x,\label{LagrangianE}
\end{equation}
where $\mathbf{v},\rho,p,\mathbf{B}$ are fluid velocity, mass density, pressure and magnetic flux density respectively, and $\gamma$ is the adiabatic index. He showed that to obtain the momentum equation as the equation of motion from this Lagrangian, the advection (continuity, adiabatic, and frozen-in) equations must be applied as constraints to the variational principle. Newcomb also showed that this constrained variational principle is equivalent to an unconstrained variational principle with this Lagrangian expressed in Lagrangian labeling $(\mathbf{x}_0,t)$. The variational constraints are no longer needed in Lagrangian labeling because the advection equations are built-in to the relabeled Lagrangian. That is, $\rho, p, \mathbf{B}$ can be expressed in Lagrangian labeling using the advection equations,
\begin{align}
\rho\,\mathrm{d}^3{x}=\rho_0\,\mathrm{d}^3x_0&\Rightarrow \rho=\rho_0/J,\label{continuity}\\
p/\rho^\gamma=p_0/\rho_0^\gamma&\Rightarrow p=p_0/J^\gamma,\label{adiabatic}\\
B_i\,\mathrm{d}S_i=B_{0i}\,\mathrm{d}S_{0i}&\Rightarrow B_i=x_{ij}B_{0j}/J,\label{frozenin}
\end{align}
where $\mathbf{x}(\mathbf{x}_0,t)$ is the configuration, $x_{ij}=\partial x_i/\partial x_{0j}$, $J=\det(x_{ij})$ is the Jacobian, and $\rho_0=\rho(\mathbf{x}_0,0)$, $p_0=p(\mathbf{x}_0,0)$, $\mathbf{B}_0=\mathbf{B}(\mathbf{x}_0,0)$. Eqs.\,(\ref{continuity})\,-\,(\ref{frozenin}) correspond to the continuity equation, adiabatic equation, and frozen-in equation respectively. Physically, they indicate that the mass enclosed in a moving volume element $\mathrm{d}^3{x}$, the specific entropy ($s=\ln(p/\rho^\gamma)$) at a moving point $\mathbf{x}(\mathbf{x}_0,t)$, and the magnetic flux through a moving area element $\mathrm{d}\mathbf{S}$ do not change as the configuration evolves. With these equations and $\mathbf{v}(\mathbf{x},t)=\dot{\mathbf{x}}(\mathbf{x}_0,t)$ substituted into Eq.\,(\ref{LagrangianE}), we obtain the Lagrangian in Lagrangian labeling,
\begin{align}
{L}(\mathbf{x},\dot{\mathbf{x}},x_{ij})=\int\bigg[ &\frac{1}{2}\rho_0\dot{{x}}^2-\frac{p_0}{(\gamma-1)J^{\gamma-1}}\nonumber\\
&-\frac{x_{ij}x_{ik}{B}_{0j}B_{0k}}{2J}\bigg]\mathrm{d}^3x_0.\label{Lagrangian3}
\end{align}
The Euler-Lagrange equation that comes from this Lagrangian is the momentum equation for ideal MHD in Lagrangian labeling,
\begin{align}
&\rho_0\ddot{x}_i-B_{0j}\frac{\partial}{\partial x_{0j}}\left(\frac{x_{ik}B_{0k}}{J}\right)\nonumber\\
&+\frac{\partial J}{\partial x_{ij}}\frac{\partial }{\partial x_{0j}}\left(\frac{p_0}{J^\gamma}+\frac{x_{kl}x_{km}B_{0l}B_{0m}}{2J^2}\right)=0,\label{momentum3}
\end{align}
which is the one and only ideal MHD equation in Lagrangian labeling. This formulation can easily be projected into 2D and 1D. It also has a few interesting variations. For instance, with the internal energy term dropped and an extra volume-preserving constraint $J=1$ added, we have a Lagrangian for incompressible MHD.

Note that a key distinction between the formulations in two labelings is the number of time-dependent variables. In Eulerian labeling there are $\mathbf{v},\rho,p,\mathbf{B}$, while in Lagrangian labeling the only one is the configuration $\mathbf{x}(\mathbf{x}_0,t)$, while $\rho_0,p_0,\mathbf{B}_0$ are time-independent parameters. This reduction of number of time-dependent variables in Lagrangian labeling is a result of the built-in advection equations. This is in contrast to Eulerian labeling where the advection equations are carried along as constraints. In the next section, we will discretize this Lagrangian in Lagrangian labeling in order to avoid a discrete constrained variational principle.

\section{discrete Lagrangian and variational integrators}\label{discretization}
Newcomb's formulation was later revisited from the perspective of geometric mechanics, and generalized into the theory of Euler-Poincar\`e reduction\cite{Holm}. From such a perspective, $\rho,s,\mathbf{B}$ are categorized as advected parameters and treated equivalently, although the advection equations look different because mass density is a 3-form, specific entropy is a 0-form, while magnetic flux density is a 2-form. 
Therefore it seems natural to respect their identities as differential forms when discretizing Newcomb's formulation. DEC\cite{Desbrun,Hirani} offers an appropriate framework for that. DEC has also been successfully applied to geometrically discretizing Lagrangians for electrodynamics\cite{Stern} and Vlasov-Maxwell systems\cite{Squire}.

DEC is a theory of differential forms on a discrete manifold, such as a simplicial complex\cite{Munkres}, i.\,e.\,a collection of simplices. In 3D, it is a tetrahedral mesh $K^3$, with the tetrahedra and their faces, edges, and vertices as 3, 2, 1, and 0-simplices respectively. A discrete $k$-form $\alpha^k$ assigns a real number to each $k$-simplex $\sigma^k$, denoted by $\langle\alpha^k,\sigma^k\rangle$, that can be interpreted as the discrete analog of $\int_{\sigma^k}\alpha^k$. Operations such as exterior derivative, wedge product and hodge star can be defined in a way that parallels their continuous definitions. For a complete treatment of DEC, see Refs.\,\onlinecite{Desbrun,Hirani}. In this paper, we will only discuss those parts of the theory that are crucial to our work.

The ideal MHD Lagrangian in Lagrangian labeling (\ref{Lagrangian3}) is not easy to discretize directly. Therefore we choose to first discretize the Lagrangian in Eulerian labeling (\ref{LagrangianE}), and then use discrete advection equations to pass into Lagrangian labeling. This same approach was adopted by Newcomb in the continuous case. In Eulerian labeling $(\sigma^k,t)$, we have a static tetrahedral mesh $K^3=\{\sigma^k\}$. The variables $\mathbf{B}$, $\rho$ and $p$ are discretized into discrete 2-form and 3-forms respectively, while $\mathbf{v}$ is discretized as a map from the vertices $\sigma^0$ to $\mathbb{R}^3$. Physically, $\mathbf{v}$ is the Eulerian velocity at the vertices. 

The first term in the Lagrangian is kinetic energy, and discretizing it involves discretizing the operation of multiplying a 3-form $\rho\,\mathrm{d}^3x$ by a 0-form ${v}^2$, evaluated as $\langle{v}^2,\sigma^0\rangle=||\mathbf{v}(\sigma^0,t)||^2$. The multiplication is discretized as follows,
\begin{equation}
\int\rho{v}^2\,\mathrm{d}^3x\rightarrow\sum_{\sigma^3}\langle\rho,\sigma^3\rangle\frac{1}{4}\sum_{\sigma^0\prec\sigma^3}\langle{v}^2,\sigma^0\rangle.\label{kinetic}
\end{equation}
The second summation is essentially averaging $\langle{v}^2,\sigma^0\rangle$ stored at the 4 vertices $\sigma^0$ of a tetrahedron $\sigma^3$. It is then multiplied with $\langle\rho,\sigma^3\rangle$ stored in this tetrahedron, and then summed over every tetrahedra in $K^3$. Note that barycentric subdivision\cite{Hirani} is implied with this discretization. We choose barycentric subdivision here because it makes averaging easier than circumcentric subdivision\cite{Hirani}.

The second term in the Lagrangian is internal energy, and its discretization is straightforward by discretizing a 3-form $p$,
\begin{equation}
\int p\,\mathrm{d}^3x\rightarrow\sum_{\sigma^3}\langle p,\sigma^3\rangle.\label{internal}
\end{equation}

The last term is magnetic energy. Mathematically, it involves the norm of a 2-form, $\mathbf{B}\cdot\mathrm{d}\mathbf{S}$. With DEC, such a norm is discretized as\cite{Desbrun,Stern}
\begin{equation}
\int {B}^2\,\mathrm{d}^3x\rightarrow\sum_{\sigma^3}\sum_{\sigma^2\prec\sigma^3}\frac{|* \sigma^2|}{| \sigma^2|}\langle B,\sigma^2\rangle^2,\label{magnetic}
\end{equation}
where $| \sigma^2|$ is the volume (area) of $\sigma^2$, and $|* \sigma^2|$ is the volume of its dual cell, namely the distance from $\sigma^2$ to the circumcenter of the tetrahedron it is a face of. Note that this norm is defined with circumcentric subdivision, because to our knowledge there is not a good discretization of such a norm with barycentric subdivision. There is no conflict between using circumcentric subdivision here and barycentric subdivision in the kinetic energy term. 

Substituting Eqs.\,(\ref{kinetic})\,-\,(\ref{magnetic}) into Eq.\,(\ref{LagrangianE}), we have a discrete Lagrangian in Eulerian labeling,
\begin{align}
L(\mathbf{v},\rho,p,\mathbf{B})=\sum_{\sigma^3}\bigg[ &\frac{\langle\rho,\sigma^3\rangle}{8}\sum_{\sigma^0\prec\sigma^3}\langle{v}^2,\sigma^0\rangle-\frac{\langle p,\sigma^3\rangle}{\gamma-1}\nonumber\\
&-\sum_{\sigma^2\prec\sigma^3}\frac{|*\sigma^2|}{2| \sigma^2|}\langle B,\sigma^2\rangle^2\bigg].\label{LagrangianED}
\end{align}
We believe there should be a constrained variational principle associated with this Lagrangian, which could lead to a variational integrator in Eulerian labeling. However, due to our current lack of understanding of discrete vector fields and Lie derivatives, we do not know how to properly discretize the variational constraints yet. 

Instead, we relabel this Lagrangian into Lagrangian labeling, where we have a moving mesh with each simplex $\sigma^k$ labeled by its origin $\sigma^k_0$. $\rho,p,\mathbf{B}$ are relabeled using the following discrete advection equations,
\begin{align}
\langle\rho,\sigma^3\rangle&=\langle\rho_0,\sigma_0^3\rangle,\label{continuityD}\\
\frac{\langle p,\sigma^3\rangle|\sigma^3|^{\gamma-1}}{\langle\rho,\sigma^3\rangle^\gamma}&=\frac{\langle p_0,\sigma_0^3\rangle|\sigma_0^3|^{\gamma-1}}{\langle\rho_0,\sigma_0^3\rangle^\gamma},\label{adiabaticD}\\
\langle B,\sigma^2\rangle&=\langle B_0,\sigma_0^2\rangle.\label{frozeninD}
\end{align}
These equations can be interpreted as discrete analogs of Eqs.\,(\ref{continuity})\,-\,(\ref{frozenin}), with $\sigma^3$, barycenter of $\sigma^3$, and $\sigma^2$ regarded as discrete analogs of volume element, point, and area element respectively. Note that if the discrete magnetic field is initially divergence-free ($\mathrm{d}B=0$), it will be guaranteed to remain so by Eq.\,(\ref{frozeninD}). Details on the discrete exterior derivative $\mathrm{d}$ can be found in Ref.\,\onlinecite{Desbrun,Hirani}.  The velocity at the vertices can be relabeled by $\mathbf{v}(\sigma^0,t)=\dot{\mathbf{x}}(\sigma_0^0,t)$, where the discrete configuration $\mathbf{x}(\sigma_0^0,t)$ stands for the position of the vertex labeled by $\sigma_0^0$. Then we can express the discrete Lagrangian in Lagrangian labeling, 
\begin{align}
L(\mathbf{x},\dot{\mathbf{x}})=\sum_{\sigma_0^3}\bigg[ &\frac{\langle\rho_0,\sigma_0^3\rangle}{8}\sum_{\sigma_0^0\prec\sigma_0^3}\dot{x}^2-\frac{\langle p_0,\sigma_0^3\rangle}{(\gamma-1)J^{\gamma-1}}\nonumber\\
&-\sum_{\sigma_0^2\prec\sigma_0^3}\frac{|*\sigma^2|}{2| \sigma^2|}\langle B_0,\sigma_0^2\rangle^2\bigg],\label{Lagrangian3D}
\end{align}
where $J=|\sigma^3|/|\sigma_0^3|$ is the discrete Jacobian. Note that $|\sigma^2|,|* \sigma^2|$ and $J$ can all be expressed in terms of $\mathbf{x}(\sigma_0^0,t)$, which makes it the only variable. There is a subtlety here, associated with the magnetic energy term discretized with circumcentric subdivision, that needs comment. As the mesh evolves, it may become not well-centered. That is, the circumcenters may move out of the tetrahedra, and $|* \sigma^2|$ will therefore become negative. But when that happens, the discretization (\ref{magnetic}) is still functional, and so is our integrator.

This Lagrangian is a geometric spatial discretization of Eq.\,(\ref{Lagrangian3}). Furthermore, by regarding its last two terms as potential energy $V(\mathbf{x}(\sigma_0^0,t))$, and rearranging the first term to be summing over vertices, the Lagrangian can be rewritten as
\begin{equation}
L(\mathbf{x},\dot{\mathbf{x}})=\sum_{\sigma_0^0}\frac{1}{2}M(\sigma_0^0)\dot{x}^2-V(\mathbf{x}),\label{Lagrangian3D2}
\end{equation}
where $M(\sigma_0^0)=\sum_{\sigma_0^3\succ\sigma_0^0}\langle\rho_0,\sigma_0^3\rangle/4$ is an effective mass for vertex $\sigma_0^0$. This Lagrangian has the form of an $N$-body Lagrangian, with $N$ being the number of vertices. The Euler-Lagrange equation following from the Lagrangian is
\begin{equation}
M(\sigma_0^0)\ddot{\mathbf{x}}=-\partial V/\partial\mathbf{x}=\mathbf{F}(\sigma_0^0).\label{momentum3D}
\end{equation}
Keep in mind that this is a spatial discretization of the MHD momentum equation in Lagrangian labeling (\ref{momentum3}).

So far, by spatial discretization, we have used a moving mesh to simulate the evolution of the fluid configuration. The spatially discretized system still has built-in advection equations and is Hamiltonian, with a conserved energy $E=\sum_{\sigma_0^0}M\dot{x}^2/2+V$. Moreover, the system is momentum conserving, in the sense that it can only gain momentum from external sources, either via forcing like gravity, or through boundaries. In our formulation, boundary conditions are applied as holonomic constraints, such as periodic boundary or rigid wall. The system cannot gain momentum from periodic boundaries. From rigid walls it can, but only in the normal direction, not the tangential directions.

Next we shall discretize the system in time in order to solve for the motion of the mesh. The built-in advection equations will be inherited after any temporal discretization. However, energy and momentum behavior is highly dependent on choice of temporal discretization. One way to ensure favorable energy and momentum behavior is to employ variational integrators\cite{Marsden,Kraus}. The idea is to temporally discretize the Lagrangian (\ref{Lagrangian3D2}) and obtain the update scheme from the discrete Euler-Lagrange equation, rather than discretizing the equation of motion (\ref{momentum3D}) directly. For example, with trapezoidal discretization, the update equation is
\begin{equation}
M(\mathbf{x}^{n+1}-2\mathbf{x}^n+\mathbf{x}^{n-1})/\tau^2=\mathbf{F}^{n}\label{update},
\end{equation}
where $n$ and $\tau$ are the number and size of the time step respectively. This update scheme is explicit and second-order accurate. In our numerical implementation, we use such a scheme as it is fast and reasonably stable. There are also other choices, such as midpoint discretization\cite{Marsden,Kraus}.

According to Refs.\,\onlinecite{Marsden,Kraus}, such schemes preserve the canonical symplectic structure on $T^*G^{N}$, the cotangent bundle of the discrete configuration space $G^{N}$, i.\,e.\,the phase space of the spatially discretized system. As $N$ becomes large, $G^{N}$ becomes ``close" to the continuous configuration space $\mathrm{Diff}(G)$, namely the diffeomorphism group\cite{Holm} on the domain $G$. And $T^*G^{N}$ becomes ``close" to the continuous phase space $T^*\mathrm{Diff}(G)$. Thus, we are preserving the canonical symplectic structure on a space that approximates the true fluid phase space. This is not the same as preserving the continuous system's symplectic structure. Yet with the symplectic structure on $T^*G^{N}$ preserved, the error of energy $E$ will be bounded in our simulations\cite{Marsden,Kraus}. Besides, a discrete Noether's theorem\cite{Marsden} states that the schemes are momentum-preserving, which means that momentum gain can only come from external sources.

Being symplectic and momentum-preserving is a major advantage of our ideal MHD integrators. However, we will not show numerical results on this in the next section, for the following two reasons. First, such properties of variational integrators have been thoroughly discussed in Ref.\,\onlinecite{Marsden}. Moreover, energy conservation does not necessarily mean the system is free of dissipation, as resistive MHD also has energy conservation. An ideal MHD algorithm can have exact energy conservation but still suffer from numerical reconnection\cite{Kraus}. 

Instead, the priority of ideal MHD simulation should be to treat the advection equations in a dissipation-free manner. After all, it is the advection of mass, entropy, and magnetic flux that defines ideal MHD. And that is exactly the second highlight of our method, which comes along with spatial discretization where discrete advection equations (\ref{continuityD})\,-\,(\ref{frozeninD}) are built-in to the spatially discretized Lagrangian (\ref{Lagrangian3D2}). The point is, we avoid error and dissipation that come with solving advection equations, now that we do not need to solve them. Such built-in advection equations are what make our schemes excel as ideal MHD integrators. In the next section, we will show results from two numerical tests that our method does not suffer from numerical reconnection, thanks to the built-in frozen-in equation.

\section{numerical results}\label{numerical}
In the previous section, all the discretization is carried out in the context of 3D compressible MHD. But just like the continuous formulation, our integrators can easily be projected to lower dimensions. Or, with an extra discrete volume-preserving constraint, we have integrators for incompressible MHD. In this section, we show results from two numerical tests with our method applied to 2D compressible MHD. The details on how the method is implemented in 2D can be found in the appendix.

\begin{figure}
\includegraphics[scale=0.6]{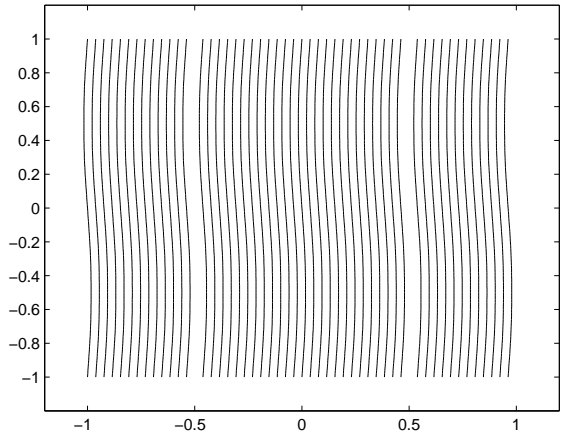}
\caption{\label{currentsheet}perturbed current sheets at $t=100$}
\end{figure}

The first test studies an equilibrium with two singular current sheets perturbed by a single mode. We borrow the setup from Ref.\,\onlinecite{Gardiner}, which is also used in Refs.\,\onlinecite{Kraus,Gawlik}. The domain is $[-1,1]\times[-1,1]$ with a resolution of $100\times100$ and periodic boundaries. The initial equilibrium is set up with $\rho=1$, $p=0.1$, $\gamma=5/3$, $B_y=1$ for $0.5<|x|\le1$, and $B_y=-1$ for $|x| \le 0.5$, and perturbed by $v_x=0.1\sin(\pi y)$. Such an equilibrium is stable in ideal MHD context, but unstable to tearing modes when finite resistivity exists. In Refs.\,\onlinecite{Gardiner,Kraus}, magnetic islands are observed to develop along the current sheets at $|x|=0.5$. In Ref.\,\onlinecite{Gawlik}, no numerical tearing is shown for the duration of the run time, which is short (till $t=4$). Our simulation is run for much longer time (till $t=100$), and shows no numerical tearing. Fig.\,\ref{currentsheet} shows the magnetic configuration at $t=100$.

\begin{figure}
\includegraphics[scale=0.6]{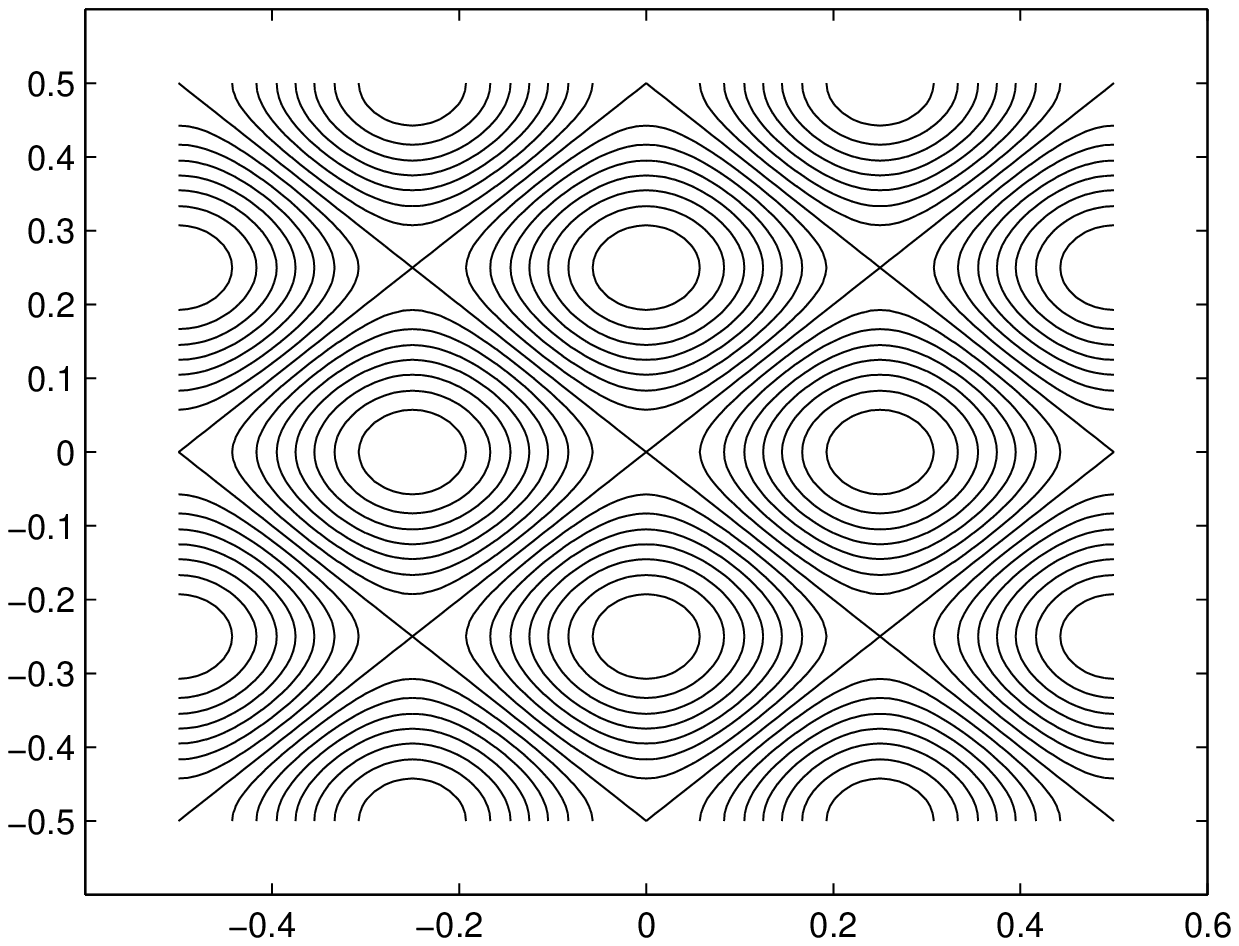}
\caption{\label{CI1}magnetic field configuration of the initial equilibrium}
\includegraphics[scale=0.6]{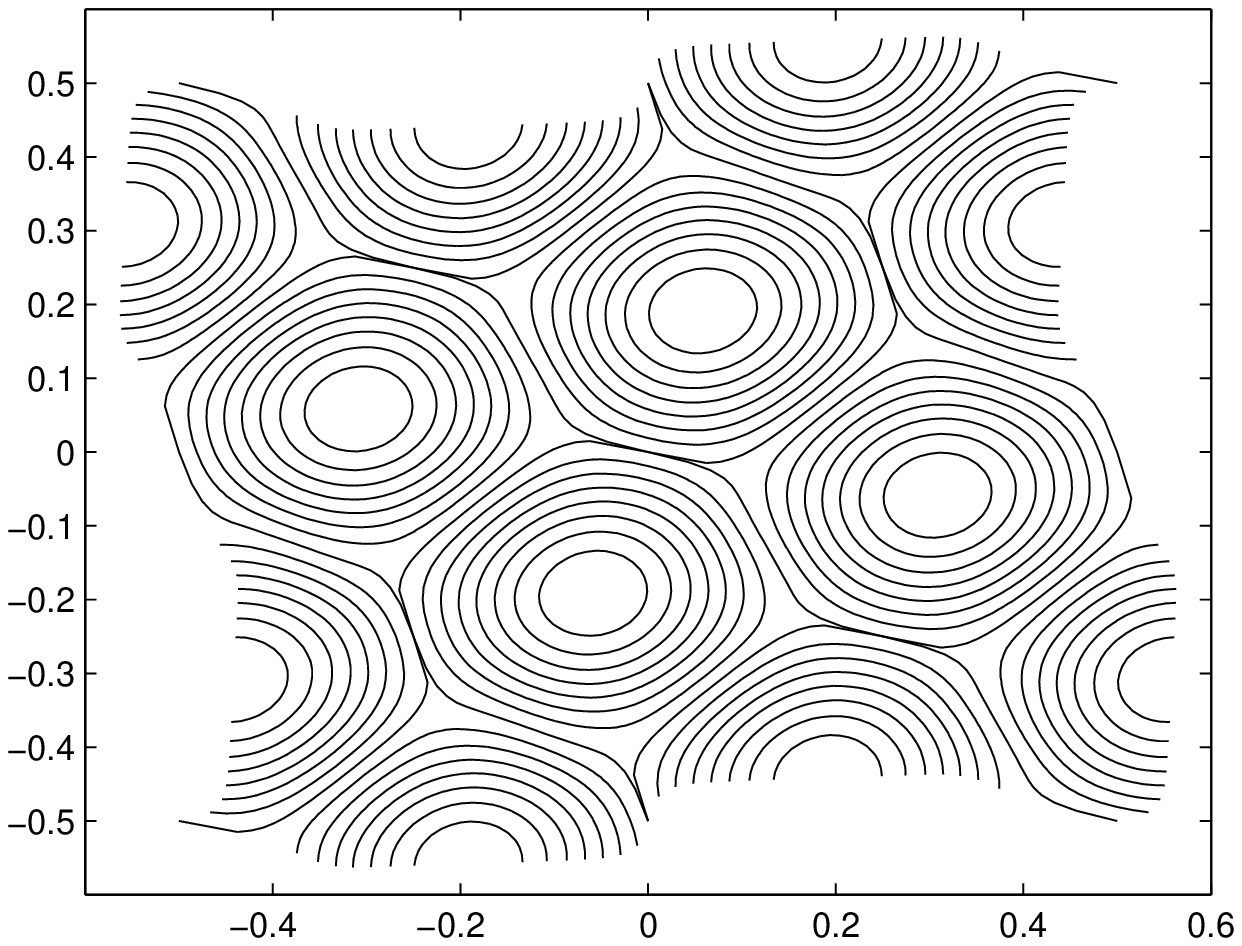}
\caption{\label{CI2}magnetic field configuration of the final equilibrium}
\end{figure}

The first test shows that our method can handle stable equilibrium with singular current sheets, which suggests that it might be best suited to studying spontaneous current sheet formation in ideal MHD \cite{Parker}. A good 2D test case for that is the coalescence instability\cite{Longcope}. The test case starts with an equilibrium with rectangular arrays of alternately twisted flux bundles, as shown in Fig.\,\ref{CI1}. According to Ref.\,\onlinecite{Parker}, this equilibrium is unstable to slipping into a close packed hexagonal array. In Ref.\,\onlinecite{Longcope}, a numerical simulation of such instability is presented, and the system is observed to reach an intermediate pentagonal equilibrium, and then the islands merge due to numerical reconnection. 

In our simulation, the domain is $[-0.5,0.5]\times[-0.5,0.5]$ with a resolution of $100\times100$ and periodic boundaries. The initial equilibrium is set up with $\rho=1$, $\mathbf{B}=\hat z\times\nabla A$ and $p=0.3+8\pi^2A^2$, where $A=0.05[\cos(4\pi x)-\cos(4\pi y)]$. We chose $\gamma=2$ so that the pressure is equivalent to the effect of an out-of-plane magnetic field, which is employed in Ref.\,\onlinecite{Longcope}. The initial perturbation is chosen to be $\mathbf{v}=0.001[\sin(2\pi x)\cos(2\pi y),-\cos(2\pi x)\sin(2\pi y)]$. It is first observed that the islands with parallel current attract each other, so the $X$-point between them gets suppressed and current builds up there. Then, when the $X$-point becomes extremely narrow and the current becomes very singular, the islands will bounce back. Since the energy error is bounded in our simulation, the islands will keep bouncing and static equilibrium will not be reached. 

To obtain an equilibrium, a friction term $-\nu\rho\dot{\mathbf{x}}$ is added to the RHS of the momentum equation in order to dissipate the kinetic energy. Our discretization of such a term follows from Ref.\,\onlinecite{Kane}. Then the system is observed to first evolve to a pentagonal structure as observed in Ref.\,\onlinecite{Longcope}, and then relax to a hexagonal equilibrium as shown in Fig.\,\ref{CI2}, which is obtained at $t=20$ with $\nu=2$. We run with multiple values of $\nu$ and the same final equilibrium is obtained. It can be seen that the $X$-points are suppressed into narrow, current-sheet-like structures in the final equilibrium. It would be interesting to distinguish whether these structures are genuinely singular current sheets, or intense but ultimately smooth current layers. However, that is beyond the scope of this paper, and we shall leave such discussion to future work.

\section{summary and discussion}
In this paper, we derive variational integrators for ideal MHD with built-in advection equations by discretizing Newcomb's Lagrangian for ideal MHD in Lagrangian labeling using DEC. The integrators possess two significant strengths. First, they are symplectic and momentum preserving, which follows from variational temporal discretization. Second, with the advection equations built-in, we avoid solving them and the accompanying error and dissipation. The latter is especially important as it allows our method to accomplish what previous methods cannot, such as handling singular current sheets without numerical reconnection. In addition, the method is physically transparent. The moving mesh practically simulates the motion of the fluid elements. It is possibly the numerical method that represents ideal MHD physics most closely.

While numerical results suggest that the method proposed here is promising, we should emphasize that it is not a panacea for all ideal MHD simulations, at least not in its current implementation. One vulnerability of the method we have recognized is that when strong shear flow is present, the simplices can get extremely deformed, and the mesh will be torn up. Such mesh distortion is a well-known problem for most numerical methods constructed on a moving mesh\cite{Springel}. Re-meshing is a popular strategy for handling it, but it appears difficult to apply it to our method in a consistent variational way.

Another issue is that presently the method can at best go only as far as ideal MHD. However, ideal MHD itself is a model with limited applicability. For example, ideal MHD fails when shocks develop. Shocks are not adiabatic and therefore finite resistivity is needed to capture it. But it is not clear to us yet how resistivity can be added to our schemes, considering the frozen-in and adiabatic equations are built-in to them.

There is one possible yet challenging way to resolve these issues, namely, by developing an Eulerian variational integrator using the discrete Lagrangian in Eulerian labeling (\ref{LagrangianED}), as discussed in Sec.\,\ref{discretization}. In that case, the mesh will be fixed, therefore mesh distortion will no longer be a problem. On the other hand, resistivity and viscosity can be added to the scheme via discrete Laplacians, which has been done successfully in Ref.\,\onlinecite{Mullen}.

Despite these issues, the strengths of the proposed method still makes it favorable for studying certain ideal MHD problems. Generally speaking, it is most suitable for problems that are shock-free, quasi-static, and with high priority on preserving the advection equations. An immediate application of this method we have in mind is to study spontaneous current sheet formation\cite{Parker}. The results from a test case in 2D, as shown in Sec.\,\ref{numerical}, suggest that it is promising. However, the problem will be much more intriguing in 3D. 

In this paper we present our formulation in ideal MHD context. Yet it can straightforwardly be generalized to other Euler-Poincar\`e fluids with advected parameters\cite{Holm}.

\begin{acknowledgements}
Y.\,Zhou would like to thank K.\,Crane, Y.\,Huang, S.\,Jardin, M.\,Kraus, C.\,Liu, Z.\,Lu, Y.\,Shi, J.\,Squire, and J.\,Stone for helpful discussions. This research was supported by the U.S. Department of Energy under Contract No.\,DE-AC02-09CH11466.
\end{acknowledgements}

\appendix
\section{numerical implementation in 2D}
\begin{figure}
\includegraphics[scale=0.35]{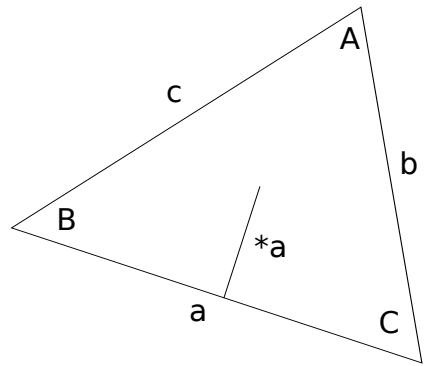}
\caption{\label{triangle}2-simplex}
\end{figure}
To project our discrete formulation to 2D, simply lower all dimensions by one. Then the mesh becomes triangular, mass density and pressure become 2-forms, and magnetic flux density becomes a 1-form. When implementing, the coefficients in the potential energy, namely $J=|\sigma^2|/|\sigma^2_0|$ and $|*\sigma^1|/|\sigma^1|$, must be explicitly expressed in terms of $\mathbf{x}(\sigma_0)$. Take the 2-simplex in Fig.\,\ref{triangle} for example, we have
\begin{align}
|\sigma^2|&=[\mathbf{x}(B)-\mathbf{x}(A)]\times[\mathbf{x}(C)-\mathbf{x}(A)]/2,\\
{|*a|}/{|a|}&=\cot(A)/2={I^a}/({2J}),
\end{align}
where $I^a=[\mathbf{x}(B)-\mathbf{x}(A)]\cdot[\mathbf{x}(C)-\mathbf{x}(A)]/(2|\sigma^2_0|)$ is defined as a normalized inner product.

\begin{figure}
\includegraphics[scale=0.35]{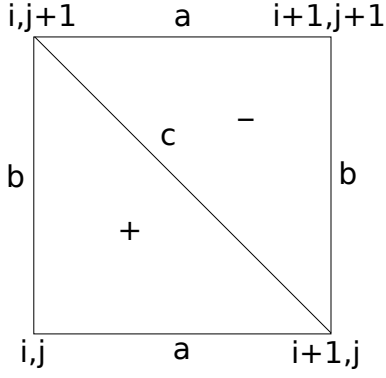}
\caption{\label{mesh}Initial mesh in 2D}
\end{figure}

So far our derivations in this paper are all carried out on an unstructured simplicial complex, on which our method should work in principle. However, unstructured meshes can involve relatively complicated data structures\cite{Elcott}. In order to simplify the data structure, we choose to start with a structured mesh when implementing. The domain is initially discretized into rectangular cells sized $s=h_ah_b$, then each rectangle is further devided into two triangles, as is shown in Fig.\,{\ref{mesh}}. In this context, the grids can be labeled with $i,j$, the lines with $i,j$ and $a,b,c$, and the triangles with $i,j$ and $\pm$. The spatially discretized Lagrangian in Lagrangian labeling (\ref{Lagrangian3D2}) then reads
\begin{align}
&L_d(x_{i,j},\dot{x}_{i,j},y_{i,j},\dot{y}_{i,j})=\nonumber\\
&\sum_{i,j}\frac{1}{2}M_{i,j}(\dot{x}_{i,j}^2+\dot{y}_{i,j}^2)-V(x_{i,j},y_{i,j}),\label{Lagrangian2d}
\end{align}
where $(x_{i,j},y_{i,j})$ is the cartesian coordinates of the grid labeled with $i,j$, and the effective grid mass $M$ is
\begin{align}
M_{i,j}=&(\rho^+_{i,j}+\rho^+_{i-1,j}+\rho^+_{i-1,j}\nonumber\\
&+\rho^-_{i-1,j}+\rho^-_{i-1,j}+\rho^-_{i-1,j-1})/3.
\end{align}
$\rho^\pm_{i,j}$ is the (initial) evaluation of $\rho$ on the triangle labeled with $i,j$ and $\pm$, and similar goes for $p^\pm_{i,j}$, $B^a_{i,j}$, $B^b_{i,j}$, and $B^c_{i,j}$. The potential energy reads
\begin{align}
V(x_{i,j},y_{i,j})=\sum_{i,j}\Bigg\{\frac{p^+_{i,j}}{(\gamma-1)(J_{i,j}^+)^{\gamma-1}}+\frac{p^-_{i,j}}{(\gamma-1)(J_{i,j}^-)^{\gamma-1}}\nonumber\\
+\frac{1}{4J^-_{i,j}}\left[\left(B^a_{i,j+1}\right)^2I^{a-}_{i,j}+\left({B^b_{i+1,j}}\right)^2{I^{b-}_{i,j}}+\left({B^c_{i,j}}\right)^2{I^{c-}_{i,j}} \right]\nonumber\\
+\frac{1}{4J^+_{i,j}}\left[\left(B^a_{i,j}\right)^2I^{a+}_{i,j}+\left({B^b_{i,j}}\right)^2{I^{b+}_{i,j}}+\left({B^c_{i,j}}\right)^2{I^{c+}_{i,j}} \right]\Bigg\},\label{potential2d}
\end{align}
where the expressions for the Jacobian $J$ are
\begin{align}
J_{i,j}^+=&[(x_{i+1,j}-x_{i,j})(y_{i,j+1}-y_{i,j})\nonumber\\
&-(y_{i+1,j}-y_{i,j})(x_{i,j+1}-x_{i,j})]/s,\\
J_{i,j}^-=&[(x_{i+1,j+1}-x_{i,j+1})(y_{i+1,j+1}-y_{i+1,j})\nonumber\\
&-(y_{i+1,j+1}-y_{i,j+1})(x_{i+1,j+1}-x_{i+1,j})]/s,
\end{align}
and the expressions for the normalized inner product $I$ are
\begin{align}
I^{a+}_{i,j}=&[(x_{i+1,j}-x_{i,j+1})(x_{i,j}-x_{i,j+1})\nonumber\\
&+(y_{i+1,j}-y_{i,j+1})(y_{i,j}-y_{i,j+1})]/s,\nonumber\\
I^{a-}_{i,j}=&[(x_{i+1,j+1}-x_{i+1,j})(x_{i,j+1}-x_{i+1,j})\nonumber\\
&+(y_{i+1,j+1}-y_{i+1,j})(y_{i,j+1}-y_{i+1,j})]/s,\nonumber\\
I^{b+}_{i,j}=&[(x_{i,j}-x_{i+1,j})(x_{i,j+1}-x_{i+1,j})\nonumber\\
&+(y_{i,j}-y_{i+1,j})(y_{i,j+1}-y_{i+1,j})]/s,\nonumber\\
I^{b-}_{i,j}=&[(x_{i+1,j}-x_{i,j+1})(x_{i+1,j+1}-x_{i,j+1})\nonumber\\
&+(y_{i+1,j}-y_{i,j+1})(y_{i+1,j+1}-y_{i,j+1})]/s,\nonumber\\
I^{c+}_{i,j}=&[(x_{i+1,j}-x_{i,j})(x_{i,j+1}-x_{i,j})\nonumber\\
&+(y_{i+1,j}-y_{i,j})(y_{i,j+1}-y_{i,j})]/s,\nonumber\\
I^{c-}_{i,j}=&[(x_{i+1,j}-x_{i+1,j+1})(x_{i,j+1}-x_{i+1,j+1})\nonumber\\
&+(y_{i+1,j}-y_{i+1,j+1})(y_{i,j+1}-y_{i+1,j+1})]/s.
\end{align}
For an update rule, we need the expressions for forces $F^x_{i,j}$ and $F^y_{i,j}$, which can be obtained by taking derivatives of the potential,
\begin{align}
F^x_{i,j}=D^{bx}_{i,j-1}-D^{bx}_{i,j}+D^{ax}_{i-1,j}-D^{ax}_{i,j}+D^{cx}_{i-1,j}-D^{cx}_{i,j-1}\nonumber\\
+R^+_{i,j}(y_{i+1,j}-y_{i,j+1})+R^+_{i-1,j}(y_{i-1,j+1}-y_{i-1,j})\nonumber\\
+R^+_{i,j-1}(y_{i,j-1}-y_{i+1,j-1})+R^-_{i-1,j-1}(y_{i-1,j}-y_{i,j-1})\nonumber\\
+R^-_{i-1,j}(y_{i,j+1}-y_{i-1,j+1})+R^-_{i,j-1}(y_{i+1,j-1}-y_{i+1,j}),\nonumber\\
F^y_{i,j}=D^{by}_{i,j-1}-D^{by}_{i,j}+D^{ay}_{i-1,j}-D^{ay}_{i,j}+D^{cy}_{i-1,j}-D^{cy}_{i,j-1}\nonumber\\
-R^+_{i,j}(x_{i+1,j}-x_{i,j+1})-R^+_{i-1,j}(x_{i-1,j+1}-x_{i-1,j})\nonumber\\
-R^+_{i,j-1}(x_{i,j-1}-x_{i+1,j-1})-R^-_{i-1,j-1}(x_{i-1,j}-x_{i,j-1})\nonumber\\
-R^-_{i-1,j}(x_{i,j+1}-x_{i-1,j+1})-R^-_{i,j-1}(x_{i+1,j-1}-x_{i+1,j}),
\end{align}
 where
 \begin{align}
R^+_{i,j}=&\frac{p^+_{i,j}}{(J^+_{i,j})^\gamma s}+\frac{1}{4(J^+_{i,j})^2s}\Big[\left(B^a_{i,j}\right)^2I^{a+}_{i,j}\nonumber\\
&+\left({B^b_{i,j}}\right)^2{I^{b+}_{i,j}}+\left({B^c_{i,j}}\right)^2{I^{c+}_{i,j}} \Big],\nonumber\\
R^-_{i,j}=&\frac{p^-_{i,j}}{(J^-_{i,j})^\gamma s}+\frac{1}{4(J^-_{i,j})^2s}\Big[\left(B^a_{i,j+1}\right)^2I^{a-}_{i,j}\nonumber\\
&+\left({B^b_{i+1,j}}\right)^2{I^{b-}_{i,j}}+\left({B^c_{i,j}}\right)^2{I^{c-}_{i,j}} \Big],
\end{align} 
and
\begin{align}
D^{ax}_{i,j}=&Q^{b+}_{i,j}(x_{i,j+1}-x_{i+1,j})+Q^{b-}_{i,j-1}(x_{i,j}-x_{i+1,j-1})\nonumber\\
&-Q^{c+}_{i,j}(x_{i,j+1}-x_{i,j})-Q^{c-}_{i,j-1}(x_{i+1,j}-x_{i+1,j-1}),\nonumber\\
D^{bx}_{i,j}=&Q^{a+}_{i,j}(x_{i+1,j}-x_{i,j+1})+Q^{a-}_{i-1,j}(x_{i,j}-x_{i-1,j+1})\nonumber\\
&-Q^{c+}_{i,j}(x_{i+1,j}-x_{i,j})-Q^{c-}_{i-1,j}(x_{i,j+1}-x_{i-1,j+1}),\nonumber\\
D^{cx}_{i,j}=&Q^{a+}_{i,j}(x_{i,j+1}-x_{i,j})+Q^{a-}_{i,j}(x_{i+1,j+1}-x_{i+1,j})\nonumber\\
&-Q^{b+}_{i,j}(x_{i+1,j}-x_{i,j})-Q^{b-}_{i,j}(x_{i+1,j+1}-x_{i,j+1}),\nonumber\\
D^{ay}_{i,j}=&Q^{b+}_{i,j}(y_{i,j+1}-y_{i+1,j})+Q^{b-}_{i,j-1}(y_{i,j}-y_{i+1,j-1})\nonumber\\
&-Q^{c+}_{i,j}(y_{i,j+1}-y_{i,j})-Q^{c-}_{i,j-1}(y_{i+1,j}-y_{i+1,j-1}),\nonumber\\
D^{by}_{i,j}=&Q^{a+}_{i,j}(y_{i+1,j}-y_{i,j+1})+Q^{a-}_{i-1,j}(y_{i,j}-y_{i-1,j+1})\nonumber\\
&-Q^{c+}_{i,j}(y_{i+1,j}-y_{i,j})-Q^{c-}_{i-1,j}(y_{i,j+1}-y_{i-1,j+1}),\nonumber\\
D^{cy}_{i,j}=&Q^{a+}_{i,j}(y_{i,j+1}-y_{i,j})+Q^{a-}_{i,j}(y_{i+1,j+1}-y_{i+1,j})\nonumber\\
&-Q^{b+}_{i,j}(y_{i+1,j}-y_{i,j})-Q^{b-}_{i,j}(y_{i+1,j+1}-y_{i,j+1}),
\end{align}
where
\begin{align}
Q^{a+}_{i,j}=\frac{(B^a_{i,j})^2}{4J^+_{i,j}s},&Q^{a-}_{i,j}=\frac{(B^a_{i,j+1})^2}{4J^-_{i,j}s},\nonumber\\
Q^{b+}_{i,j}=\frac{(B^b_{i,j})^2}{4J^+_{i,j}s},&Q^{b-}_{i,j}=\frac{(B^b_{i+1,j})^2}{4J^-_{i,j}s},\nonumber\\
Q^{c+}_{i,j}=\frac{(B^c_{i,j})^2}{4J^+_{i,j}s},&Q^{c-}_{i,j}=\frac{(B^c_{i,j})^2}{4J^-_{i,j}s}.
\end{align}

Boundary conditions are applied as holonomic constraints. For example, periodic boundary in $x$ can be realized with $x_{N_x,j}=x_{1,j}+L_x$, and rigid wall with $x_{1,j}=0,x_{N_x,j}=L_x$. $N_x$ and $L_x$ stand for the resolution and domain size in $x$ respectively.


%
%

%



\end{document}